\documentclass[twocolumn,aps,rmp,floats]{revtex4}
\usepackage{graphics,graphicx,epsfig}
\usepackage{amssymb}
\usepackage{epsf,epstopdf,wrapfig}

\begin{document}

\title{Synergy from Silence in a Combinatorial Neural Code}

\author{Elad Schneidman,$^{a-c}$ Jason L. Puchalla,$^a$ Ronen Segev,$^a$  Robert A. Harris,$^a$ William Bialek$^{b,c}$  and Michael J. Berry II$^a$}

\affiliation{$^a$Department of Molecular Biology, 
$^b$Joseph Henry Laboratories of Physics, and
$^c$Lewis--Sigler Institute for Integrative Genomics,
Princeton University, Princeton, New Jersey 08544 USA}

\date{\today}

\begin{abstract}
The manner in which groups of neurons represent events in the external world is fundamental to neuroscience.  Here, we analyze the population code of the retina during naturalistic stimulation and show that the information conveyed by specific multi-neuronal firing patterns can be very different from the sum of the patternÕs parts. Synchronous spikes convey more information than either of the participating cells, but almost always less than their sum, making them redundant coding symbols.  Surprisingly, patterns of spiking and silence are mostly synergistic Ð carrying information that is accessible only by observing the whole pattern of activity, rather than its components, and signifying unique features in the stimulus. These results demonstrate that the retina uses a combinatorial code and that the brain can benefit significantly from recognizing multi-neuronal firing patterns.
\end{abstract}

\maketitle

While much of our understanding of the brain is derived from studies of single neurons, it is clear that neural systems generally rely on populations of neurons to represent stimuli and to direct motor outputs.   In particular, concerted firing among groups of neurons has been observed in many different neural systems, often occurring more frequently than would be expected if neurons were acting independently \cite{Perkel+al_67,meister+al_95,Vaadia+al_95}.  However, many different views of how concerted firing patterns represent information collectively have been proposed, even within the same brain region \cite{meister+al_95,Vaadia+al_95,Abbott+Dayan_99,Zohary+al_94,dan+al_98,Shamir+Sompolinsky_04,Nirenberg+al_01,Schneidman+al_03b,Puchalla+al_05}.  In one view, the unreliability of single neurons requires that many neurons encode the same information, and only by averaging over large groups can encoded information be extracted reliably \cite{Shadlen+Newsome_94}. Another hypothesis is that the information conveyed by multi-neuronal spiking patterns is simply the sum of information that each of the cells conveys on its own; this view is related to ideas of redundancy reduction \cite{Barlow_61}, decorrelation \cite{Atick_92} and independent component analysis \cite{Bell+Sejnowski_97,vanHateren+vanderScaaf_98}, which have been suggested as design principles for efficient coding.  Finally, there is the possibility of concerted coding \cite{Meister_96}, in which synchronous multi-neuronal activity patterns can be synergistic, conveying more information than the sum of contributions from the individual neurons. Such a coding scheme has the capacity to multiplex many distinct messages onto the axons of a small number of neurons.

To assess these different coding schemes, the average amount of information that small groups of cells convey about natural and artificial stimuli has been compared to the sum of single cell contributions.   This analysis has been carried out in many systems Ð vertebrate retina \cite{Puchalla+al_05}, primary visual cortex \cite{Gawne+Richmond_93,Reich+al_01}, primary somatosensory cortex \cite{Petersen+al_01}, the auditory pathway \cite{Chechik+al_01} and motor cortex \cite{Gat+Tishby_99,Narayanan+al_05}. Typically, the deviation from independence has been small, and redundancy is most common.  However, synergistic response patterns may simply be rare events that have only a small effect on the average mutual information.  In particular, synchronized spiking has been shown to be correlated with specific stimulus features or motor outputs \cite{meister+al_95,Vaadia+al_95,Abeles_91,Riehle+al_97,hatsopolous+al_98}, and has therefore been suggested as a special symbol in the population code of many neural circuits.

These issues can be addressed by focusing on the information conveyed by {\em specific} multi-cell activity patterns \cite{Brenner+al_00} rather than estimating the mutual information between stimuli and responses, which is an {\em average} over all multi-cell patterns \cite{Schneidman+al_03b}. This is a natural framework in which to analyze a population code, because specific firing patterns are the actual neural symbols that downstream circuits encounter in real time.  Using this analysis, we find significant synergy and redundancy in different patterns of ganglion cell firing.  Averaged over patterns, these contributions often partially cancel, explaining the relatively low values commonly observed in previous studies. We study the stimulus features that correspond to multi-neuronal firing patterns and relate synergy to the functional properties of individual neurons.  Finally, we show that similar properties emerge from a generic model of neural responses, suggesting that combinatorial coding may exist in other neural circuits, too.

\section*{Results}
Simultaneous recordings were made from many ganglion cells in the salamander and guinea pig retinas using a multi-electrode array, while artificial and natural movies were presented on a computer monitor (see Methods).  Movie clips 20-30 s long were repeated many times (100-150) in order to sample both the diversity and variability of ganglion cell firing patterns.  An example of the spike trains recorded from two cells is shown in Fig. 1A.

\subsection*{Individual Neural Symbols} The amount of information that the arrival time of a neural ÔsymbolÕ Ð such as a spike or any other specific pattern of neuronal activity in an ensemble (see below) Ð conveys about a complex, time-varying stimulus S is given by \cite{Brenner+al_00}:

\begin{equation}
I(\sigma;S) = \frac{1}{T} \int_0^T \frac{r_{\sigma}(t)}{\bar{r}_{\sigma}} \log_2 \frac{r_{\sigma}(t)}{\bar{r}_{\sigma}} dt
\end{equation}   

\noindent where $\sigma$  is the neural symbol, $r_{\sigma}(t)$  is the time dependent rate of the symbolÕs occurrence, and $\bar{r}_{\sigma}$  is the time average of   over the entire duration of the stimulus $T$.  The symbol rate $r_{\sigma}(t)$  measures the probability per unit time of the symbol occurring, and  is estimated in the same way as we estimate time-dependent spike rates, by forming a peri-stimulus time histogram (PSTH) using repeated trials of the stimulus. As with the conventional time-dependent firing rate for single spikes, modulations in this symbol rate reflect reproducible locking of the events  $\sigma$  to dynamic features in the stimulus.  Neuronal symbols that occur with a rate that is strongly modulated by the stimulus convey more information per symbol than those with a weakly modulated rate. Notice that equation 1 depends only on the variation in the time-varying symbol rate, $r_{\sigma}(t)$, compared with the average symbol rate, $\bar{r}_{\sigma}$. Using this relationship, we estimated the information carried by different ganglion cell firing patterns (see Methods).

\begin{figure}
\begin{centering}
\vskip -0.5 in
\centerline{\psfig{figure= 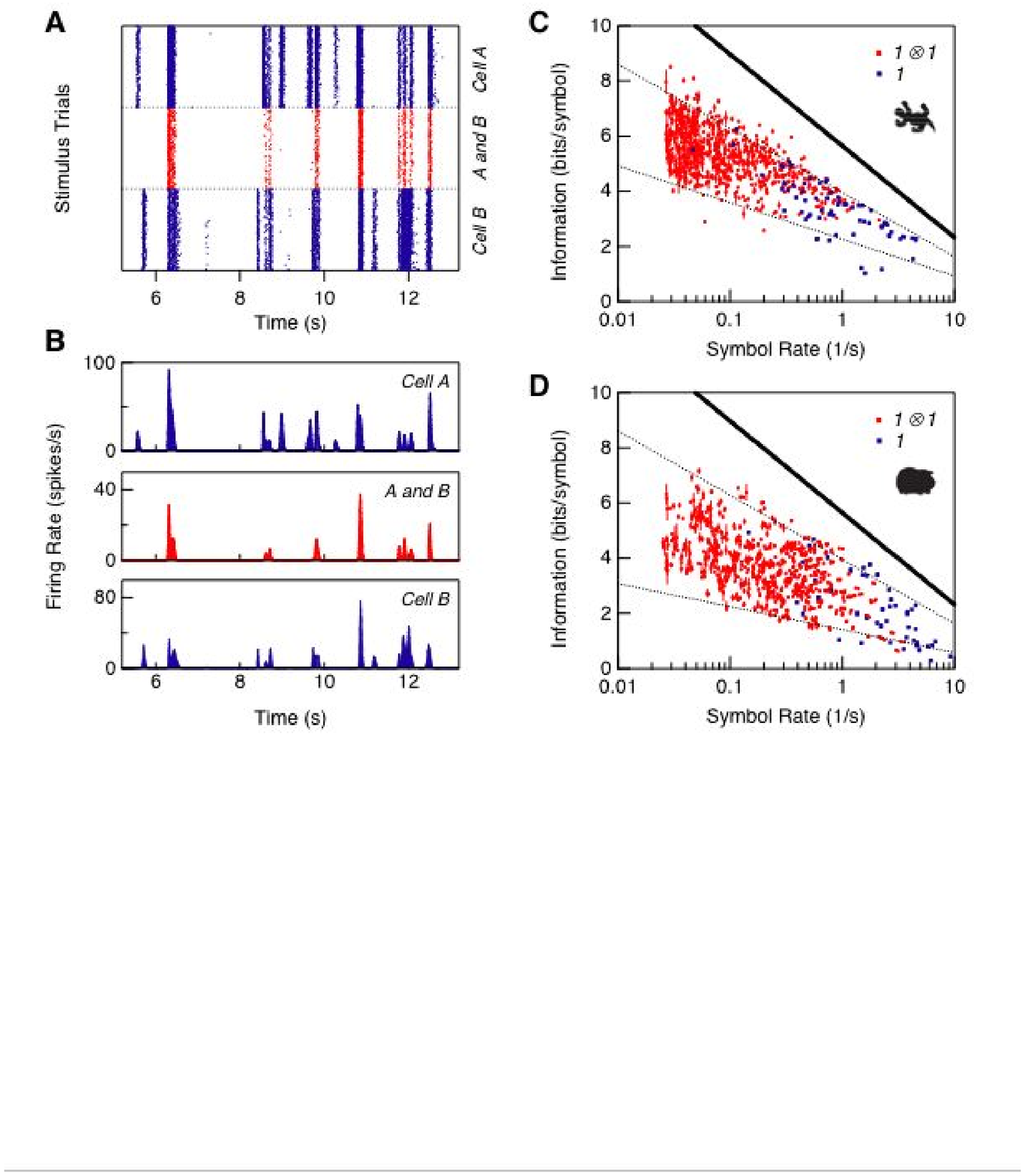,width=1.05\columnwidth}}
\vskip  -1.4 in
\caption{
Information carried by single spikes and synchronous spike pairs. {\bf A}. Typical response segment of ganglion cells to repeated presentations of a natural movie. Top and bottom panels show the spikes of two single cells (blue); middle panel shows the events of synchronized spiking of these cells within $\pm$10 ms (red).  {\bf B}. Post stimulus time histograms of the spike rasters in panel A. {\bf C, D}. Information conveyed by single spikes (blue; every dot stands for one ganglion cell) and synchronized spike pairs (red; every dot stand for one cell pair) as a function of symbol rate for the (salamander, guinea pig).  Data points shown from 2 different natural movies. Solid line marks the upper bound for information transmission by a symbol at the given rate with a 20 ms temporal precision [28]; dashed lines are 40$\%$ and 70$\%$ of this maximum for the salamander and 25$\%$/70$\%$ for the guinea pig.}
\end{centering}
\end{figure}

The simplest neural symbol is the single spike, denoted as $Ô1Õ$. Over the whole set of natural movies, the average information conveyed by a single spike in the salamander was  $\langle I (1_A;S)\rangle_{A} = 3.6 \pm 1.0$ bits (mean $\pm$ standard deviation across the population; n = 204 cells) and in the guinea pig was 2.2 $\pm$ 1.2 bits (n=59 cells).  No clear dependence on the particular movie clip was observed.
 
For synchronous spiking among pairs of neurons, we defined the compound symbol  $1_A \otimes 1_B$ denoting all the spikes of cell A that had a spike of cell B within $\pm$ 10 ms (see Fig. 1A).  This narrow temporal window for synchrony matches the time scale over which excess synchronous pairs have been observed \cite{meister+al_95} and also corresponds with the typical duration of excitatory post-synaptic currents evoked by AMPA receptors \cite{Pouille+Scanziani_04}.

The information that synchronous spiking conveyed about natural movies was significantly higher than for single spikes,  $\langle I(1_A\otimes1_B; S) \rangle_{\{A,B\}} = 5.3 \pm 1.0$ bits (Fig 1C; n=2887 pairs).  In particular, for all cell pairs the symbol $1_A \otimes 1_B$  was at least as informative, and usually much more informative than either  $1_A$ or $1_B$  alone (data not shown).  Changing the temporal window used to define synchrony between $\pm$5 ms and $\pm$50 ms had a negligible effect on the symbolÕs information.  Interestingly, the  $1 \otimes 1$ symbols obeyed a similar relation to that of the single spikes: given our time resolution of 20 ms, nearly all symbols conveyed an information between 40$\%$ and 70$\%$ of the maximum possible information determined by that symbolÕs average event rate (Fig. 1C) \cite{spikes}.  Similar behavior was found for the guinea pig (Fig. 1D), but with somewhat higher firing rates (4.2 $\pm$ 1.2 Hz vs. 1.2 $\pm$ 0.8 Hz) and lower coding efficiency ($33\% \pm  5 \%$ vs. $55\% \pm 11\%$).

\subsection*{Combinatorial Codes} 
To analyze the existence and nature of combinatorial codes in the retina, we compared the information transmitted by a compound symbol Ð a multiÐneuronal firing pattern Ð to the sum of information values conveyed by the components of the compound symbol.  In the case of two neurons, we define:
\begin{equation}
\Delta I(\sigma_A \otimes \sigma_B;S) = I(\sigma_A \otimes \sigma_B;S) - I(\sigma_A;S) - I(\sigma_B;S)
\end{equation} 

If $\Delta I$  is zero, then the components of the compound symbol contribute independent (additive) information about the visual stimulus. Otherwise, a combinatorial code exists, and downstream circuits can benefit by interpreting multi-neuronal firing patterns as distinct coding symbols.  In particular, synergy $(\Delta I > 0)$ implies that the compound symbol encodes information that is not available at the level of single neuron events.

\subsubsection*{Synchronous Spiking}
For natural movies, we found that synchronous spikes were significantly less informative than expected from summing the information contributed by each individual spike.  The redundancy of synchronous spikes ranged up to  $\Delta I = -4.1$ bits in salamander (-47$\%$ of the constituent symbol information) and -2.7 bits in guinea pig (-22$\%$).  For cells spaced less than 200 $\mu m$ apart, which have overlapping receptive field centers,  $\Delta I = -1.1 \pm 0.84$ bits (n=2887 pairs) for salamander and -0.41 $\pm$ 0.72 bits (n=766 pairs) for guinea pig.  No clear differences were found between movie clips (n=9 clips).  Synchronized firing tended to produce more negative $\Delta I$  values between nearby cells than faraway cells (Fig 2A). We found a small fraction of cell pairs with synergy for synchronous spike pairs; in almost all cases, one cell was ON-type and the other was OFF-type (data not shown).

\begin{figure}
\begin{centering}
\vskip -0.6 in
\centerline{\psfig{figure= 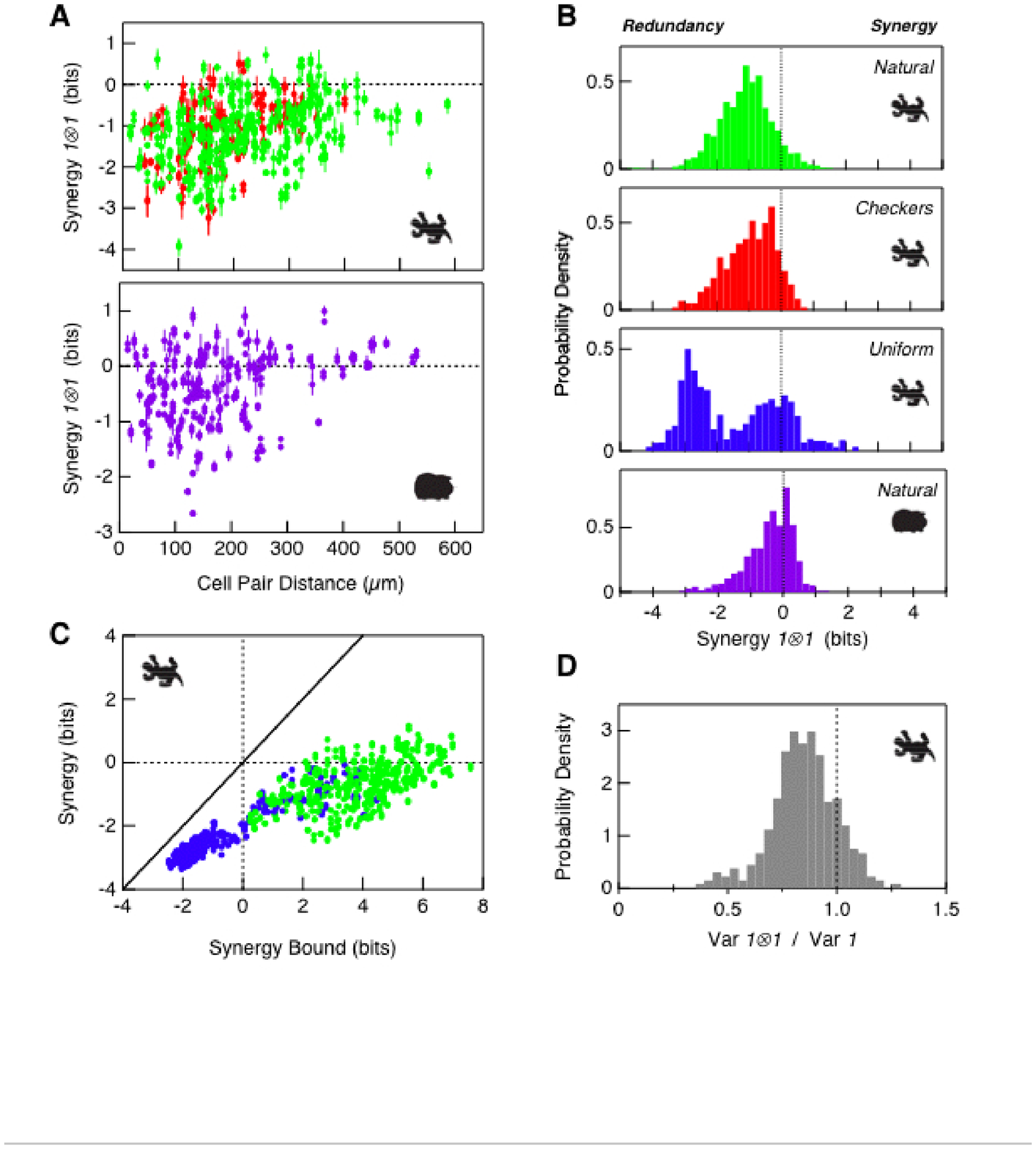,width=1.05\columnwidth}}
\vskip  -0.5 in
\caption{Synergy of synchronous spike pairs.  {\bf A}. Synergy of synchronous spike pairs, $1 \otimes 1$,  as a function of the distance between the receptive field center of these cells (every dot stands for one cell pair). Above: for salamander, natural movie data is collected over 3 different movies (green); checkerboard movie data is collected over 2 different movies (red).  Below: for guinea pig, natural movie data is collected over 1 movie (purple). {\bf B}.  Distribution of synergies for $1 \otimes 1$  symbols in salamander for natural movies (green, 2556 cell pairs), checkerboard  movies (red, 382 pairs) and uniform flicker (blue, 749 pairs); synergies for $1 \otimes 1$  symbols in guinea pig for natural movies (purple, 766 pairs).  {\bf C}.  Actual  synergy for $1 \otimes 1$  symbols plotted versus the upper bound on their synergy derived from the PSTH of each cell (see Methods) for natural movies (green) and uniform flicker (blue). Error bars omitted for clarity.  {\bf D}. Distribution of stimulus certainty for $1 \otimes 1$  symbols divided by the certainty for single spikes.  Certainty is defined as the variance of the symbol-triggered stimulus distribution (see text).
}
\end{centering}
\end{figure}

To further interpret these results, we carried out a similar analysis for artificial stimuli.  In spatially uniform flicker, where all ganglion cells see the same input, the largest values of redundancy were similar to those found with naturalistic stimulation, but many more cell pairs had high values of redundancy (Fig. 2B).  This makes sense because the strong spatial correlations in natural images will cause ganglion cells with highly overlapping receptive fields to experience roughly, but not exactly, the same visual input. In checkerboard flicker, where spatial correlations only extend up to the size of square regions (55 $\mu m$), synchronous spike pairs were less redundant,  $\Delta I = -0.94 \pm 0.89$ bits for cells spaced by less than 200 $\mu m$ (Fig. 2B, n=567 pairs).  Together these results suggest that the degree of redundancy for synchronous spike pairs is largely determined by the properties of the stimulus.

The simple correlation structure of artificial movies also allows us to relate neural events to the visual stimulus \cite{dRvS+Bialek_88}. Using spatially uniform flicker, we typically found that the symbol-triggered average stimulus (STA) for $1 \otimes 1$  symbols was intermediate between the STAs of its component spikes.  However, the stimuli preceding a synchronous spike pair clustered more tightly around the STA than for single spikes.  Figure 2D shows that the variance along the direction in stimulus space defined by the STA was consistently smaller for $1 \otimes 1$  symbols than for single spikes (see Methods). These results indicate that the extra information conveyed by $1 \otimes 1$ symbols compared to single spikes is due, in part, to increased certainty that the stimulus is similar to the STA.

The fact that most of the $1 \otimes 1$  symbols were less informative than their sum needs to be put in context: how much information can a synchronous spike pair possibly convey? Intuitively, if two neurons each are very precisely locked to the stimulus, then there is little scope for synchronous spikes to say anything different from that said by the single neurons.  We explored the potential for synergy between two neurons by using an iterative algorithm for finding the most informative synchronous spike train given the firing rate of each individual neuron (see Methods).  Fig. 2C compares the potential synergy for    $1 \otimes $ symbols with their actual value. Under naturalistic stimulation, cell pairs had considerable room for synergy but instead were mostly redundant (green).   Under spatially uniform flicker, we could not always find a possible combination of spikes into synchronous pairs that was synergistic (blue), but all synchronous spike pairs we significantly more redundant than they could have been.

\subsubsection*{Spiking and Silence}
In a combinatorial code, the combination of spiking in some neurons and silence (non-spiking) in other neurons can also be an important event.  Silence alone is not highly informative: the information content of the symbol Ô0Õ is 0.011 $\pm $ 0.015  bits for 20 ms of silence and 0.074 $\pm$ 0.079 bits for 100 ms of silence (natural movies, salamander, n=204 cells). However, the symbol $1_A \otimes 0+B$  , where cell A fires a spike and cell B does not spike within 50 ms of cell AÕs spike, often was found to be synergistic $( \Delta I > 0)$ both for natural and for artificial movies (Fig. 3A).  Importantly, for every ganglion cell, there is another cell whose silence adds synergistically to the spiking cell (Fig. 3A).  Synergy from silence tended to increase when the temporal window used to define silence increased; we chose 50 ms because this corresponds to the typical duration of ISPSs \cite{Pouille+Scanziani_04,Gibson+al_99}, consistent with a likely mechanism for detecting combinations of spiking and silence (see below, Fig. 6).

\begin{figure}
\begin{centering}
\vskip -0.75 in
\centerline{ \hskip 1.3 in \psfig{ figure= 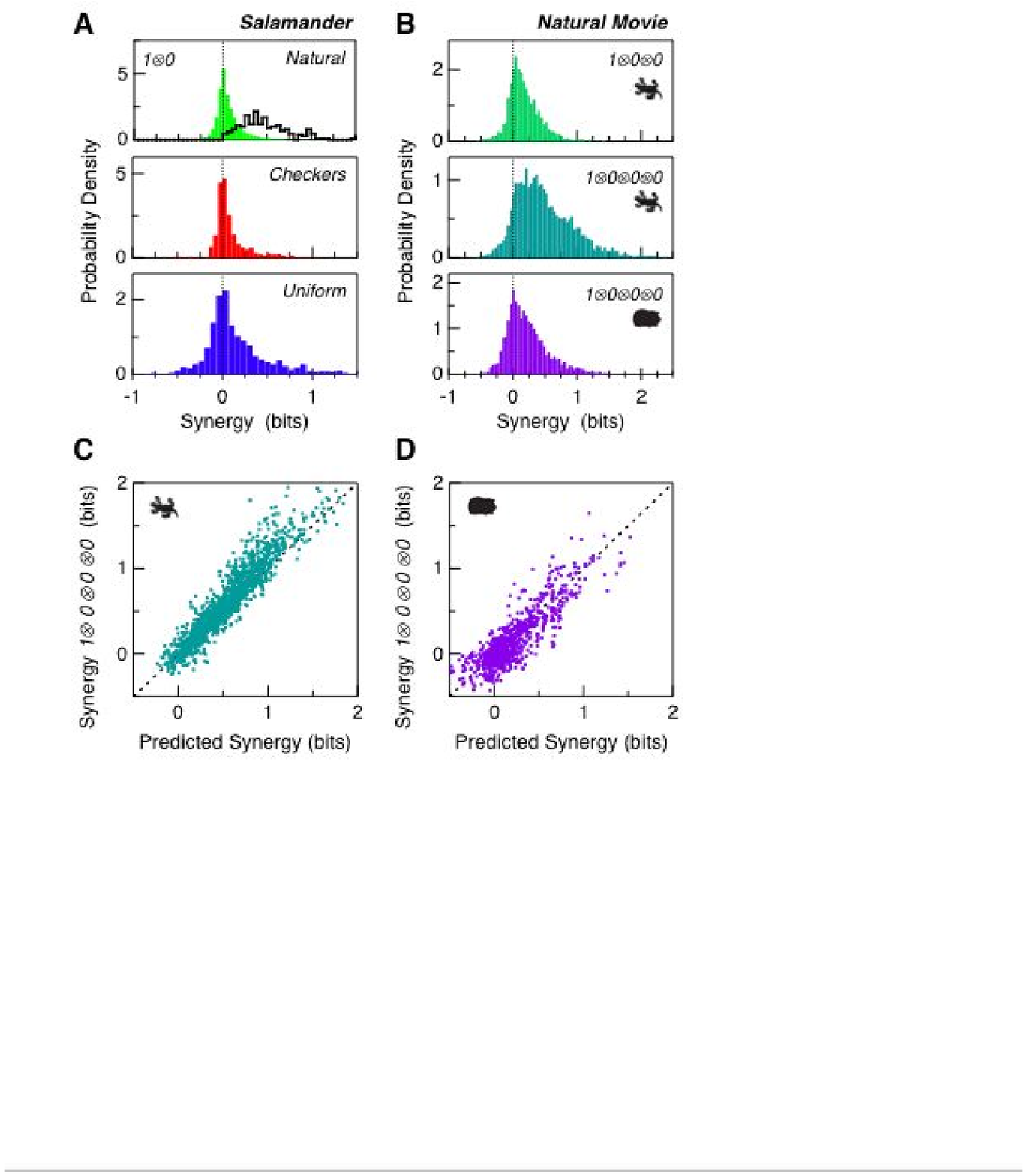,width=1.5\columnwidth}}
\vskip  -2.05 in
\caption{Synergy from silence.  {\bf A}.  Distribution of synergy values of the $1 \otimes 0$  symbol for natural movies (top panel, 5497 cell pairs), checkerboard movies (middle panel, 688 pairs) and uniform flicker (683 pairs). Top panel also shows the distribution of maximum  $1 \otimes 0$ synergies for each cell found by pairing it with its most informative silent partner (black).  {\bf B}.  Distribution of synergy values of $1 \otimes 0 \otimes 0$   (top) and $1 \otimes 0 \otimes 0 \otimes 0$  symbols (middle) for salamander and  $1 \otimes 0 \otimes 0 \otimes 0$   for guinea pig (bottom panel) under naturalistic stimulation. {\bf C,D}.  Synergy of  $1 \otimes 0 \otimes 0 \otimes 0$ symbols compared to the sum of synergy values of the corresponding three synergy of $1 \otimes 0$  symbols for (salamander, guinea pig) under naturalistic stimulation.  Error bars omitted for clarity.
}
\end{centering}
\end{figure}

Synergy from silence becomes even more apparent for firing patterns with one spiking cell and several silent cells: for natural movies,  $1 \otimes 0 \otimes 0$ symbols were more synergistic than $1 \otimes 0$ , and $1 \otimes 0 \otimes 0 \otimes 0$  symbols were even more so, with synergy exceeding 2 bits in some cases (Fig 3B).  Here, synergy from silence is a generalization of Eqn. 2: for instance, $\Delta I(1\otimes 0 \otimes 0; S) = I(1 \otimes 0 \otimes 0; S) - I(1;S) - I(0 \otimes 0; S)$. The synergy for   symbols was 0.50 $\pm$ 0.46 bits (n=6438 quadruplets) for salamander and 0.23 $\pm$ 0.33 bits (n=2455 quadruplets) for guinea pig. The synergistic contribution from the silence of 3 neurons is roughly equal to the sum of the synergies from the silence of each neuron combined pairwise with the spiking neuron (Figs. 3C,D).

To understand why the combination of spiking and silence can be synergistic, it is instructive to look at examples of such coding symbols under random flicker stimulation. Synergy results when the requirement of silence from the second neuron (cell B) effectively vetoes a subset of the firing events produced by the spiking cell (cell A), while leaving other events essentially unchanged (Fig. 4A).  Because this vetoing operation is stimulus specific, the STA of the $1 \otimes 0$  symbol is qualitatively different from the STA of cell A alone (Fig. 4B).  In fact, the visual message represented by a  $1 \otimes 0$ symbol was often different from {\em any} of the STAs of individual neurons (Fig. 4B).  Thus, many distinct visual messages are represented by combinations of spiking and silence in the population of retinal ganglion cells.

\begin{figure}
\begin{centering}
\vskip -0.6 in
\centerline{\psfig{figure= 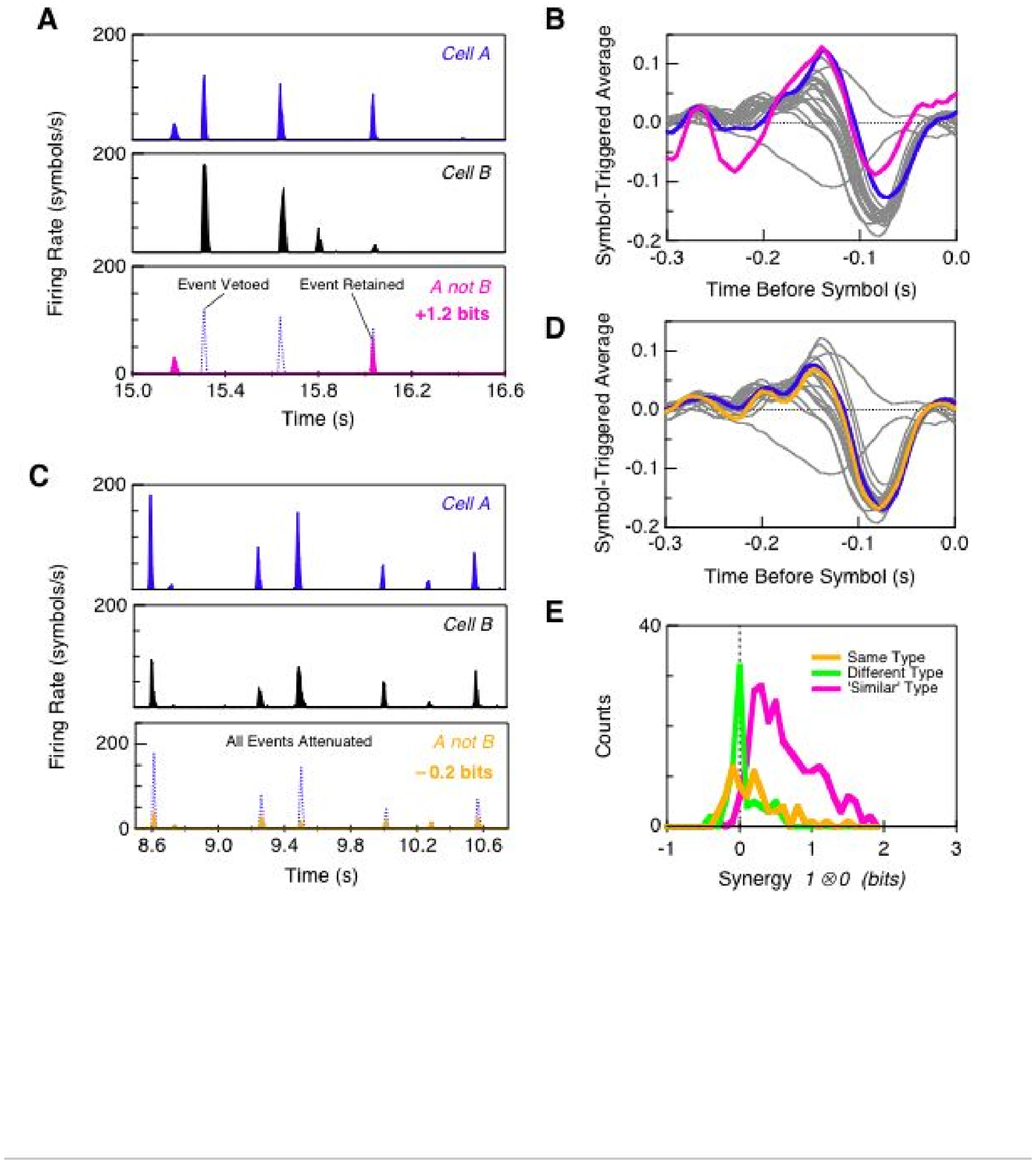,width=1.05\columnwidth}}
\vskip  -0.9 in
\caption{Origins of synergy from silence.  {\bf A}.  Example of information gain from silence.  Firing rate of two cells is shown (cell A, blue; cell B, black) along with the combination of cell A firing and cell B silent (bottom, pink).  {\bf B}.  Symbol-triggered average for cell A (blue) and A not B (pink) against the other 20 cells recorded under uniform flicker (grey). {\bf C}.  Example of information loss from silence.  Firing rate of two cells is shown (cell A, blue; cell B, black) along with the combination of cell A firing and cell B silent (bottom, orange).  {\bf D}.  Symbol-triggered average for cell A (blue) and A not B (orange) against the other 20 cells recorded under uniform flicker (grey).  {\bf E}.  Histogram of  $1 \otimes 0$ synergies for cell pairs of the same function type (orange), different broad type (green), and similar functional type (pink).  See Methods for details of functional classification.}
\end{centering}
\end{figure}

Synergy from silence depends critically on the identity of the two neurons: if they are functionally very similar, then the requirement of silence from cell B indiscriminately attenuates all of the firing events produced by cell A (Fig. 4C).  Now, the  $1 \otimes 0$ symbol encodes the same visual message as the single spike (Fig. 4D), but this compound symbol has many fewer events than either individual spike.  In fact, this combination can even result in a loss of information relative to that conveyed by a single spike (Fig. 4C).  Information loss occurs when the difference between the most sharply-locked firing events and the typical events decreases, as can be seen in Fig. 4C.  These results demonstrate that synergy from silence is not a generic consequence of the decreased event rate for  $1 \otimes 0$ symbols relative to single spikes, but instead depends on the detailed pattern of firing exhibited by the compound symbol.

The lack of synergy from silence for functionally similar neurons was found to be systematic: the values of $\Delta I$  for cells of the same functional subtype (see Methods) were found to be scattered close to zero (Fig. 4E, orange).  Interestingly, the symbol-triggered stimulus variance for $1 \otimes 0$  symbols formed from cells of the same functional type was slightly but significantly greater than the stimulus variance for single spikes: variance ratio = 1.06 $\pm$ 0.018 (S.E.M., n=71 pairs).  This result is the opposite of what was found for synchronous spike and indicates the information loss here results, in part, from decreased certainty that the stimulus is similar to the STA.  We found also that  $1 \otimes 0$  symbols formed from cells with very different function, such as pairs of ON and OFF cells, were not synergistic (Fig. 4E, green).  This is expected, because such cells have no signal correlation and no noise correlation \cite{Puchalla+al_05,Segev+al_06}. Instead, $1 \otimes 0$ symbols formed from similar but different neurons Ð same broad functional type (e.g., fast OFF) but different subtype \cite{Schneidman+al_03a} Ð usually were synergistic under spatially uniform stimulation (Fig. 4E, pink).

\subsection*{Model of a Combinatorial Code}
To better understand the origin of combinatorial coding, we constructed a simple model of retinal processing.  We assumed that ganglion cell light responses were described by what is called the LN model, where the visual stimulus is first filtered by the classical receptive field and then passed through a sigmoidal threshold function that truncates negative values to produce a time-varying firing rate (see Methods).  Multiple ganglion cells had different receptive fields and were assumed to be correlated only through their receptive field overlap (also known as conditional independence or lack of noise correlation).

For this model of ganglion cell light responses, the classical receptive field can be thought of as a vector in the high dimensional space of all possible visual stimuli (Fig. 5A).  The firing rate of each neuron depends only on the overlap between the stimulus and its receptive field, which can be described by an angle $\theta$  between the stimulus and the neuronÕs Ôpreferred directionÕ in stimulus space (Fig. 5B). This overlap angle summarizes both the spatial and temporal integration of the classical receptive field.  We choose the simple case of a uniform distribution of overlap angles for the population of ganglion cells, allowing us to calculate the information that single cells, pairs and triplets convey about the visual stimulus.  

\begin{figure}
\begin{centering}
\vskip -0.4 in
\centerline{\psfig{figure= 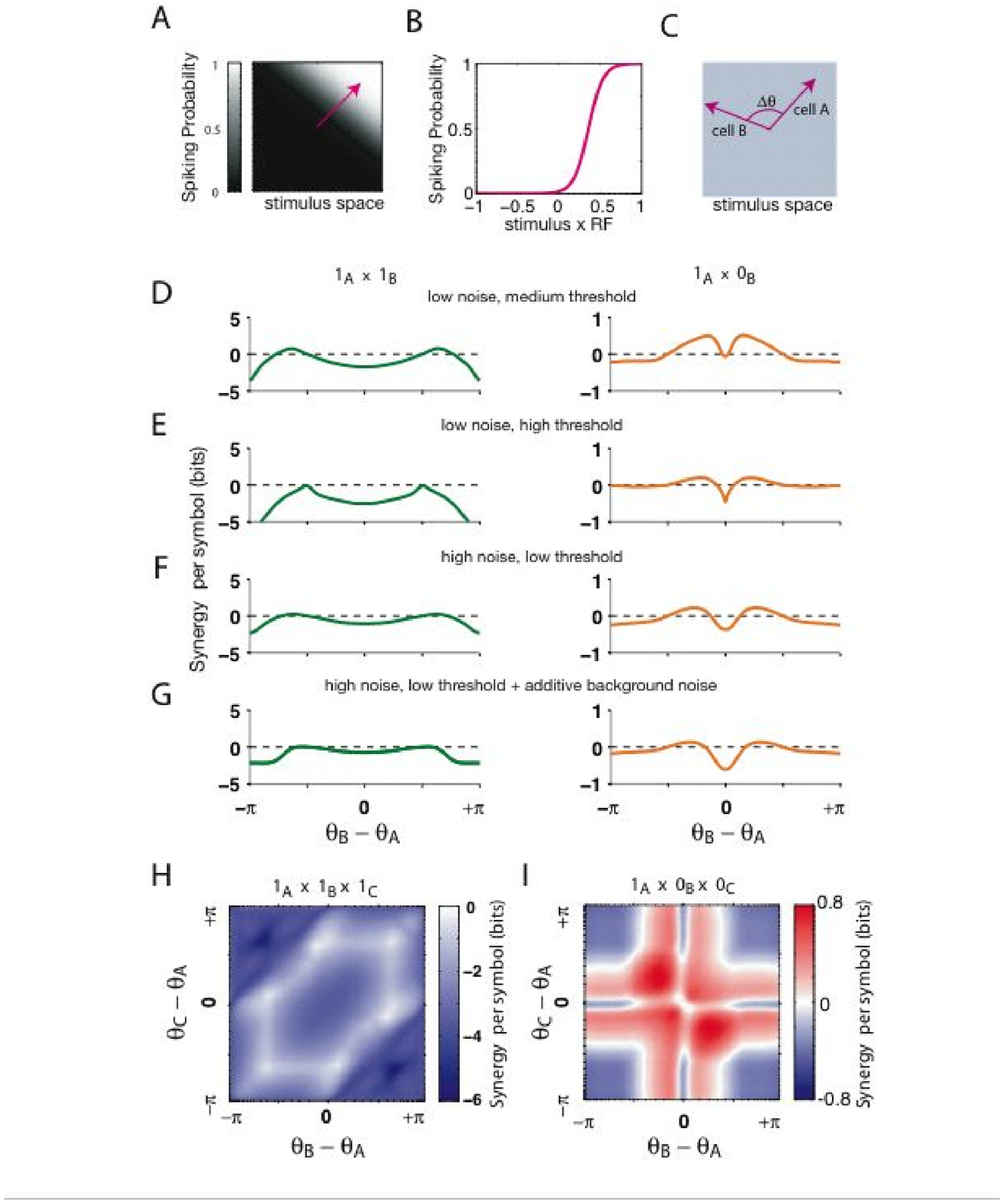,width=1.25\columnwidth}}
\vskip  -0. in
\caption{Synergy per symbol in a simple neural population model.  {\bf A}. Stimuli are vectors in the space of all possible stimuli, here depicted in two-dimensions.  The preferred stimulus of a neuron is described by the direction of a unit vector in stimulus-space (pink arrow), defining a spiking probability along this direction (grey scale). All cells in the model are identical except for their preferred direction. {\bf B}. The spiking probability of the neuron from panel A shown as a function of the overlap between the stimulus and the cellÕs preferred direction, $\theta$. {\bf C}. For two cells, the symmetry of the model implies that the joint information is only a function of the angle between the preferred directions of the cells, $\Delta \theta$. {\bf D}. Synergy for $1 \otimes 1$  symbols (left column) and $1 \otimes 0$  symbols (right column) shown as a function of the angle between the cellsÕ preferred stimuli, $\Delta \theta$. Model parameters are $\beta=8$, $\alpha=5$, without additive background noise $\zeta=0$ (see Methods). {\bf E}. Same as in D, but with a higher spiking threshold $(\alpha=8)$, which results in more selective spiking. {\bf F}. Same as in D, but with model parameters $\beta=4$, $\alpha=2$, and $\zeta=0$, corresponding to a lower selectivity and a noisier response. {\bf G}. Same as in F, but with an additive noise $\zeta=0.02$.  {\bf H}. Synergy for the triplet symbol $1 \otimes 1 \otimes 1$  shown as a function of the angle between cell 1 and cells 2 (x-axis), and the angle between cells 2 and cell 3 (y-axis). Same model parameters as in D. {\bf I}. Same model as H, but for the triplet symbol  $1 \otimes 0 \otimes 0$ .}
\end{centering}
\end{figure}

Similar to our experimental results, we found that  $1 \otimes 1$ symbols tend to be either redundant or nearly independent (Fig. 5D, left).  On the other hand,  $1 \otimes 0$ symbols had many synergistic combinations and some weakly redundant ones (Fig. 5D, right).  This results closely resembles the pattern of synergy from silence for different cell types: cells with nearly identical function $(\Delta \theta \sim 0)$ were weakly redundant, ON/OFF pairs (large $\Delta \theta$) were independent, and cells that were similar but different (intermediate $\Delta \theta$) almost all exhibited synergy from silence.  Changing the neuronsÕ selectivity or noise level (see Methods) gave qualitatively similar results (Fig. 5E-G).  For triplets, we found that synchronized spiking $1 \otimes 1 \otimes 1$ was always redundant, whereas $1\otimes 0 \otimes 0$  patterns demonstrated strong synergistic combinations and some weakly redundant ones (Fig. 5H,I) Ð in striking resemblance to the results for real ganglion cells. We emphasize that this model does not include stimulus dependent correlations between cells (also called Ònoise correlationsÓ), showing that the combinatorial coding can result from stimulus-induced correlations alone.

\section*{Discussion}
A combinatorial retinal code that relies on synergy from silence has great flexibility. Because most ganglion cells respond sparsely to natural stimuli, synchronous spike pairs are relatively rare (Fig. 1C,D), but combinations of spiking and silence are quite common.  Because the receptive fields of ganglion cells overlap extensively Ð covering visual space $\sim$60 times over \cite{Segev+al_04} Ð a single neuron can form combinatorial firing patterns with many nearby ganglion cells. How can the brain select interesting population symbols to recognize?  Since almost all  $1\otimes 0 \otimes 0 \otimes 0$ symbols are synergistic, such symbols could simply be chosen at random.  However, if combinations of spiking and silence from ganglion cells of the same functional type were avoided, the brain could significantly enhance the chance that remaining combinations were highly synergistic.  Generic learning rules, such as spike-timing dependent plasticity, may also allow the brain to focus in further on the most informative retinal firing patterns.

How might the central brain recognize combinations of spiking and silence?  We propose that they are detected by a simple circuit involving feedforward inhibition along with monosynaptic excitation (Fig. 6). Networks of inhibitory interneurons clearly allow cortical circuits to synchronize and oscillate \cite{Beierlein+al_00,Traub+al_98}, but interneurons possess extreme functional diversity \cite{Gupta+al_00,Parra+al_98} suggesting more detailed roles in information processing than currently appreciated \cite{Beierlein+al_03,Miles_00}. In particular, many subtypes exhibit very fast and reliable transmission with relatively little spike frequency adaptation or synaptic depression \cite{Gibson+al_99,Martina+al_00}.  These properties are ideal for vetoing excitation \cite{Swadlow_03,Gabernet+al_05}, thereby detecting combinations of spiking and silence in the input to a cortical circuit.

\begin{figure}
\begin{centering}
\vskip -0.7 in
\centerline{\psfig{figure= 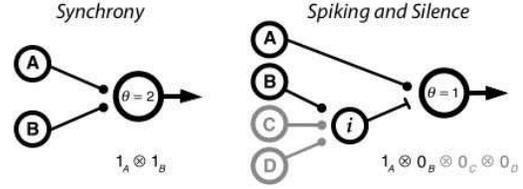,width=1.1\columnwidth}}
\vskip  -2.9 in
\caption{Schematic of circuit for recognizing compound symbols. {\em Left}: Cells A and B have excitatory synapses onto a readout cell with a threshold of two simultaneous inputs.  {\em Right}: Spiking cell (A) has a direct excitatory synapse onto a readout neuron.  Silent cell (B) excites an inhibitory cell (i), which feeds forward onto the readout neuron with a threshold of one spike.  When cell B spikes, this pathway indirectly vetoes excitation of the readout neuron; as a result, the readout neuron will only be excited when A fires but B is silent.  This network is easily generalized to confer sensitivity to silence from multiple neurons by having all such neurons excite the same inhibitory cell (cells C, D).  Such feedforward inhibitory circuitry has been described in the cortex \cite{Gibson+al_99}.}
\end{centering}
\end{figure}

The combinatorial code of the retina combines elements of all three hypotheses about population codes, but in ways that are different than previously thought \cite{Meister_96}. Redundancy was found in the synchronous events.  At the same time, synchronous spikes encoded more information than either constituent spike alone.  Some of this additional information resulted from increased certainty about the visual stimulus \cite{dRvS+Bialek_88,Gabbiani+al_96} Ð a form of averaging over spikes to achieve higher fidelity.  Ganglion cells of different broad functional types were found to be largely independent.  Synergy was found in combinations of spiking and silence, rather than in synchronous spikes. Why might the retinaÕs population code be structured in this complex fashion?  Redundancy can benefit a neural code by increasing the robustness of representation and by tagging patterns of neural activity to be learned \cite{Barlow_01}.  At the same time, synergistic compound symbols constitute a huge vocabulary of different visual features multiplexed onto optic nerve fibers and represented at the fastest response time of the circuit.

The simple model we presented suggests that combinatorial coding does not require elaborate circuit mechanisms or strong noise correlations, nor is it highly specific to the processing carried out in the retina.  This implies that similar results may be found in other neural circuits.  Furthermore, average synergy and redundancy have been reported for cortical neurons \cite{Petersen+al_01,Gat+Tishby_99,Narayanan+al_05}, suggesting that these circuits can also signal with synergistic and redundant ensemble firing patterns. Because the same neurons can participate in both synergistic and redundant firing patterns whose contributions largely cancel, even the observation of average independence \cite{Gawne+Richmond_93,Reich+al_01} does not preclude combinatorial coding.  Similar results have been found for spike patterns produced by a single neuron, where spike pairs separated by short time intervals were synergistic but pairs separated by longer time intervals were redundant \cite{Brenner+al_00}.  As the methods we have used to analyze the retinal code can readily be extended to other neural systems, this approach promises to yield new insights into the nature of population neural codes throughout the brain.

\section*{Methods}

{\bf Electrophysiological Recording.}  Eyes were dissected from the salamander or guinea pig and cut into pieces that left the retina and pigment epithelium intact.  These pieces were placed with the ganglion cells facing a multi-electrode array and were perfused with oxygenated RingerÕs at room temperature (salamander) or AmesÕ medium at $36^\circ$C (guinea pig). Stable recordings of over 12 hours were achieved under these conditions. Extracellular voltages were recorded by a MultiChannel Systems MEA 60 microelectrode array and streamed to disk for offline analysis.  Spike waveforms were sorted using the spike size and shape in a 2.5 ms window.  Only well isolated spike waveforms were used: we required that spike trains had fewer than 0.5 $\%$ of the inter-spike interval less than 2 ms and that fewer than 1 $\%$ of the spikes were found within $\pm$ 0.2 ms of spikes from other cells. This study is based on measurements of 8210 (902) cell pairs between 369 (61) cells recorded from 5 (4) retinas in the salamander (guinea pig) with multiple movie clips.  There were between 9 and 40 cells recorded per movie clip; some of the same cells were recorded during multiple clips.

{\bf Visual Stimulation.} Natural movie clips were acquired using a Canon Optura Pi video camera at 30 frames per second.  Movies were taken of woodland scenes and included several qualitatively different kinds of motion: objects moving in a scene, optic flow, and simulated saccades.  Checkerboard flicker consisted of 55 $\mu m$ square regions on the retina that were randomly chosen to be either black or white every 33 ms.  In spatially uniform flicker, light intensities were chosen randomly every 33 ms from a Gaussian distribution with a standard deviation equal to 18 $\%$ of the mean. All visual stimuli were displayed on an NEC FP1370 monitor and projected onto the retina using standard optics.  The mean light level was $12 mW/m^2$ at the retina.

{\bf Functional classification.}  Salamander ganglion cells were divided into 3 broad classes using the time course of their spike-triggered average (STA) as well as their responses to diffuse steps of light.  Fast OFF cells (~75$\%$ of population) had short latency responses to both the onset and offset of light.  Slow OFF cells (~10 $\%$) had longer latency responses and only responded at the offset of light.  ON cells ($\sim$15$\%$) only responded at the onset of light.  These definitions correspond to those previously made using random flicker stimulation Ð Ôfast OFFÕ, Ôweak OFFÕ, ÔONÕ \cite{Schnitzer-Meister_03} Ð as well as using flashes Ð ÔON/OFFÕ, ÔOFFÕ, ÔONÕ \cite{Burkhardt +al_98}.  Fast OFF cells were resolved into 6 subtypes using an information theoretic method of functional classification \cite{Schneidman+al_03a}; this method is sensitive to the entire response function of a neuron, rather than just its spike-triggered average.  Cells of the same subtype had nearly identical STAs; different subtypes had characteristic shifts in their response latency as well as the degree to which their STAs were biphasic versus monophasic.

{\bf Information carried by a neural symbol.}   We first estimated the time-dependent rate at which each neural symbol $\sigma$ occurs, $r_{\sigma}(t)$ , by binning the spike train in time windows $\Delta t$   and counting the number of spikes in each window across all stimulus trials.  We then used Eqn. 1 to estimate the information that symbol $\sigma$ conveys about the stimulus with $\Delta t$  as a free parameter.  We extrapolated to the limit $\Delta t \rightarrow 0$ , giving a bin-free estimate of the information \cite{Brenner+al_00}. To correct for undersampling and bias, a jackknife approach was used, in which we estimated the information in subsets of the data and extrapolated to an infinite number of repeats \cite{strong+al_98}. We have omitted information values of symbols when the symbol rate was too low to give a reliable estimate of the information.  An alternative approach  measures the information conveyed by individual occurrences of an event \cite{DeWeese+Meister_99}.  Our definition instead measures the average information conveyed by the specific time at which a single event occurs.  As discussed in \cite{Brenner+al_00}, these two measures are consistent, and Eqn. 1 can be derived in many ways, including as a limit of the information per event defined in \cite{DeWeese+Meister_99}.  We emphasize that Eqn. 1 is an exact measure of the information carried by the arrival time of a single event and makes no assumptions about the correlations between events.  In Fig 2, we defined stimulus certainty by convolving the stimulus with the symbol-triggered average, compiling the distribution of such values whenever the symbol occurred, and calculating the variance of that distribution.

{\bf Potential for synergy.}  We estimated how synergistic the joint spiking symbol $1 \otimes 1$  could have been, given the firing rates of its constituent cells, $r_{1A}(t)$  and $r_{1B}(t)$ . At every time point, the $1 \otimes 1$  symbol rate could have been anywhere between the minimal value $r_{AB}^{min}(t) = max \{ 0 \, ,\, r_{1A}(t)+r_{1B}(t)-1/\Delta t \}$, where $\Delta t$ is size of the time bin, and the maximal value  $r_{AB}^{max}(t) = min \{r_{1A}(t) \, , \, r_{1B}(t)\}$. Because information per symbol is higher for larger modulations of the symbol rate, the maximum information will be obtained if at each time the rate takes either its maximal or minimal value; different mean symbol rates can be obtained by distributing different time points between their maximal and minimal symbol rates. We start from  $r_{AB}^{min}(t)$ and add $1/N\Delta t$ to the rate of one of the time bins (where $N$ is number of repeated presentation of the stimulus), which corresponds to adding one synchronous spike pair during all $N$ stimulus trials.  We choose the time bin such that this addition results in the largest increase in information that would be carried by the $1 \otimes 1$  symbol. Then, we seek pairs of bins for which raising the rate in one bin and reducing the rate in the other bin by the same amount (i.e. keeping $\bar{r}_{AB}$  constant) would further increase the information.  We stop when no such pair is found, thus achieving a locally optimal solution for distributing synchronous spike pairs across time to achieve the maximal possible information. We repeat this procedure for each possible value of $\bar{r}_{AB}$ (in steps of $1 / N \Delta T$ ) and find the largest $1 \otimes 1$  information over all the possible values of $\bar{r}_{AB}$. The potential for synergy in the $1 \otimes 1$  symbol is given by the difference between this maximum possible information for the  $1 \otimes 1$ symbol and each cellÕs individual information. We note that our greedy algorithm gives a lower bound on the synergy that the two cellÕs firing rates could achieve. 

{\bf Model of ganglion cell light responses.}  Each ganglion cell has a receptive field given by a unit vector $\vec{v}_i$  in the space of all visual stimuli.  This preferred direction has an overlap with all stimuli $\vec{s}$  defined by the angle $\theta_i = \vec{v}_i\cdot \vec{s}$.  The spiking probability of each cell in response to a stimulus, $P(1_i|\vec{s})$, is given by a sigmoidal function $\frac{1}{2}\left[ 1+\tanh(\beta(\vec{v}_i \cdot \vec{s}) -\alpha)\right]$  , where $\alpha$ is a bias term, controlling the threshold of the neuron, and $\beta$  is the slope of the sigmoid, controlling how deterministic the cellÕs response is (the firing probability will be either zero or one with a very sharp threshold). We added more noise in the form of a background firing rate, implemented by adding $\zeta$  to the spiking probability (and truncating to the range $\left[0,1\right]$).  To estimate the synergy or redundancy of pairs and triplets of cells, we averaged responses over a uniformly distributed stimulus ensemble (randomly sampled with 50,000 stimuli) and a set of 100 neurons with preferred directions in stimulus space ranging from 0 to $2 \pi$ in equal jumps.  The joint spiking probability of cells A and B is equal to  $P(1_A|\vec{s})P(1_B|\vec{s})$. The information per symbol values and synergy values were calculated as for the retinal ganglion cells.

\acknowledgments {We thank Markus Meister and Adrienne Fairhall for critical discussions of the manuscript, and Nadiya Tkachuk for her help with the guinea pig retina experiments. This work was supported a grant from the National Eye Institute (EY-014196) and a Pew Scholarship (M.J.B.).
Correspondence and requests for materials should be addressed to ES or MJB: \{elads,berry\}@princeton.edu.
}

\end{document}